\documentclass[a4paper]{article}
\usepackage{graphicx}
\usepackage{amsmath}
\usepackage{bm}		% enables bold italic in math mode

\newcommand{\beq}{\begin{equation}}
\newcommand{\eeq}{\end{equation}}
\newcommand{\dif}[1]{\mathrm{d} #1 \,}

\newcommand{\GeVc}{GeV/$c$}
\newcommand{\GeVcc}{GeV/$c^2$}
\newcommand{\rarr}{\rightarrow}
\newcommand{\phiH}{\phi_\mathrm{h}}
\newcommand{\phiS}{\phi_\mathrm{S}}
\newcommand{\kT}{\vec{k_\mathrm{T}}}
\newcommand{\kTsq}{k_\mathrm{T}^2}

\newcommand{\PhT}{\vec{P_\mathrm{T}}}
\newcommand{\PhTscal}{P_\mathrm{T}}

\newcommand{\qT}{\vec{q_\mathrm{T}}}
\newcommand{\qTscal}{q_\mathrm{T}}
\newcommand{\qTscalt}{\texorpdfstring{$q_\mathrm{T}$}{q\_T}}
\newcommand{\kaT}{\vec{k_{\pi\mathrm{T}}}}

\newcommand{\kbT}{\vec{k_{N\mathrm{T}}}}

\newcommand{\fSiv}[1]{f_{1\mathrm{T}}^{\perp #1}}

\newcommand{\fSivp}[1]{f_{1\mathrm{T},\mathrm{p}}^{\perp #1}}
\newcommand{\F}[2]{F_\mathrm{#1}^{#2}}
\newcommand{\A}[2]{A_\mathrm{#1}^{#2}}
\newcommand{\qqswap}{(q\leftrightarrow\bar{q})}

\renewcommand{\vec}[1]{\bm{#1}}

\newcommand{\ack}{\section*{Acknowledgements}}

\usepackage[pdftex]{hyperref}

\hypersetup{
  pdftitle={Measurement of q_T-weighted TSAs in 2015 COMPASS Drell-Yan data},%
  pdfauthor={J. Matousek, on behalf of the COMPASS collaboration},%
  pdfsubject={},%
  pdfkeywords={COMPASS, fixed-target, nucleon structure, 
				transverse spin asymmetries, SIDIS, Drell-Yan, weighted},%
  pdfstartview={},%
  bookmarksopen=true, breaklinks=true, debug=true, %
  colorlinks=true, linkcolor=blue, citecolor=blue, urlcolor=blue
}
\begin{document}
\title{Measurement of \qTscalt-weighted TSAs in 2015 COMPASS Drell--Yan data}

\author{Jan Matou\v{s}ek\footnote{
			Charles University, Faculty of Mathematics and Physics,
				180\,00 Prague, Czech Republic
			and Trieste Section of INFN, 341\,27 Trieste, Italy.
			\href{mailto:jan.matousek@cern.ch}{jan.matousek@cern.ch}.
		}, on Behalf of the COMPASS Collaboration}
\date{}

\maketitle
%%%%%%%%%%%%%%%%%%%%%%%%%%%%%%%%%%%%%%%%%%%%%%%%%%%%%%%%%%%%%%%%%%%%%%%%%%%%%%%
\begin{abstract}
In the polarised Drell--Yan experiment at the COMPASS facility at CERN the beam of negatively-charged pions with 190\,GeV/$c$ momentum and intensity about $10^8$~pions/s interacted with transversely polarised NH$_3$ target. Muon pairs produced in Drell--Yan process (DY) were detected. Recently, the first ever Transverse Spin Asymmetries (TSAs) measurement in DY has been presented by COMPASS. A complementary analysis of the TSAs weighted by powers of the dimuon transverse momentum $q_\mathrm{T}$ are presented. In the Transverse Momentum Dependent (TMD) PDF formalism, the $q_\mathrm{T}$-weighted TSAs can be written in terms of products of the TMD PDFs of two colliding hadrons, unlike the conventional TSAs, which are  their convolutions over quarks transverse momenta. The results are compared in a straightforward way with the weighted Sivers asymmetry in the SIDIS process, released by COMPASS in 2016.
\end{abstract}
%%%%%%%%%%%%%%%%%%%%%%%%%%%%%%%%%%%%%%%%%%%%%%%%%%%%%%%%%%%%%%%%%%%%%%%%%%%%%%%
\section{Introduction}

The hadron structure can be described at leading twist by eight Transverse Momentum Dependent (TMD) Parton Distribution Functions (PDFs), which depend on the fraction $x$ of the hadron momentum carried by the parton and the transverse component of the parton momentum $\kT^2$. They have been probed in Semi-Inclusive Deep Inelastic Scattering (SIDIS), where the cross-section contains convolutions of the TMD PDFs and fragmentation functions\,\cite{bacchetta:2007}.

Comparison of SIDIS with the Drell--Yan process (DY), giving access to convolutions of TMD PDFs of the two colliding hadrons\,\cite{arnold:2008}, can provide a test of the PDFs universality. In fact, two PDFs (the Sivers and Boer--Mulders functions) are predicted to bear opposite signs when extracted from SIDIS and DY\,\cite{collins:2002}. Recently, COMPASS has done a pioneering measurement of the Transverse Spin Asymmetries (TSAs) in DY\,\cite{compass:2017dy}. A complementary analysis of the same data, using the formalism of transverse momentum \qTscalt\ weighted TSAs\,\cite{efremov:2004, sissakian:2005b}, is presented.

The convolutions of TMDs are usually solved assuming a certain functional form of their dependence on $\kT^2$ (e.g. Gaussian). In SIDIS, it can be avoided using the TSAs weighted with powers of the outgoing hadron transverse momentum~$\PhTscal$\,\cite{kotzinian:1996, kotzinian:1997, boer:1998}. Preliminary $\PhTscal$-weighted TSAs from HERMES\,\cite{gregor:2005} were used to estimate the \qTscalt-weighted Sivers asymmetry expected in DY experiments\,\cite{efremov:2004}. Similarly, in Sec.~\ref{sec:projection} we use the recent COMPASS measurement of the $\PhTscal/z$-weighted Sivers asymmetry\,\cite{bradamante:2017} to get a projection, which we compare with our DY results.
	
%%%%%%%%%%%%%%%%%%%%%%%%%%%%%%%%%%%%%%%%%%%%%%%%%%%%%%%%%%%%%%%%%%%%%%%%%%%%%%%
\section{Transverse momentum weighted asymmetries in Drell--Yan process}
\label{sec:measurement}
%%%%%%%%%%%%%%%%%%%%%%%%%%%%%%%%%%%%%%%%%%%%%%%%%%%%%%%%%%%%%%%%%%%%%%%%%%%%%%%
We study the Drell--Yan reaction with 190\,\GeVc\ pion beam and NH$_3$ target with the H nuclei transversely polarised 
$\pi^- \mathrm{p}^\uparrow \rarr \mu^- \mu^+ X.$
At Leading Order (LO), the reaction proceeds via annihilation of a quark-antiquark pair into a virtual photon with momentum $q$, which decays into the dimuon. The LO cross-section contains five orthogonal modulations in $\phi$ and $\phiS$ -- the azimuthal angles of the muon momentum in the Collins--Soper frame and of the target spin vector $\vec{S_\mathrm{T}}$ in the target rest frame with $z$-axis along the beam momentum and $x$-axis along dimuon transverse momentum $\qT$, respectively\,\cite{compass:prop2}. The structure functions $F_\mathrm{U,T}^X$ can be written as convolutions of TMD PDFs over the intrinsic transverse momenta of the two colliding partons $\kaT$ and $\kbT$\,\cite{arnold:2008}. When the structure functions are integrated over \qTscalt\ with properly chosen weights, the convolutions can be disentangled\footnote{We use the same Sivers function sign as in Ref.\,\cite{bradamante:2017, martin:2017} and opposite to Ref.~\cite{bacchetta:2007,arnold:2008} and the Trento convention\,\cite{trento}.%, i.e. the same as\,\cite{bradamante:2017,martin:2017} and opposite to\,\cite{bacchetta:2007,arnold:2008,compass:prop2}. Under this convention, there is minus sign in \refeq{eq:dy_ftphis_w} and plus sign in \refeq{eq:wSivSIDISdef}.
}:
\begin{align}
	\label{eq:dy_fu1_w}
		\int \dif{^2\qT} \F{U}{1}
		&= \hphantom{-} \frac{1}{3} \sum_q e_q^2 
					\bigl[	f_{1,\pi}^{\bar{q}}(x_\pi) \, 
							f_{1,N}^q(x_N) + 
							f_{1,\pi}^{q}(x_\pi) \, 
							f_{1,N}^{\bar{q}}(x_N)\bigr] \\
%	\label{eq:dy_fu2phi_w}
%		\int \dif{^2\qT} \frac{\qTscal^2}{4 M_\pi M_\mathrm{p}} \F{U}{\cos2\phi}
%		&= \hphantom{-}\frac{2}{3} \sum_q e_q^2 
%			\bigl[  h_{1,\pi}^{\perp (1) \bar{q}} (x_\pi)\,
%					h_{1,N}^{\perp (1) q} (x_N)
%					+ \qqswap \bigr],	\\
	\label{eq:dy_ftphis_w}
		\int \dif{^2\qT}\frac{\qTscal}{M_\mathrm{p}}\F{T}{\sin\phiS}
		&= - \frac{2}{3} \sum_q e_q^2 \bigl[
				f_{1,\pi}^{\bar{q}}(x_\pi) \, 
				\fSivp{(1)q}(x_N) + \qqswap \bigr],	\\
	\label{eq:dy_ftphip_w}
		\int \dif{^2\qT} \frac{\qTscal^3}{2 M_\pi M_\mathrm{p}^2} 
				\F{T}{\sin(2\phi+\phiS)}
		&= - \frac{2}{3} \sum_q e_q^2 \bigl[
				h_{1,\pi}^{\perp (1) \bar{q}} (x_\pi) \,
				h_{1\mathrm{T},\mathrm{p}}^{\perp (2) q} (x_N)
				+ \qqswap \bigr], \\
	\label{eq:dy_ftphim_w}
		\int \dif{^2\qT} \frac{\qTscal}{M_\pi} \F{T}{\sin(2\phi-\phiS)}
		&= - \frac{2}{3} \sum_q e_q^2 \bigl[
				h_{1,\pi}^{\perp (1) \bar{q}} (x_\pi) \, h_{1,\mathrm{p}}^q (x_N)
				+ \qqswap \bigr],
\end{align}
where the sums run over quarks and antiquarks $q$; $e_q$ are fractional electric charges; $M_{\pi,\mathrm{p}}$ are the pion and proton masses; and $f^{(n)}$ or $h^{(n)}$ are the $n$-th $\kTsq$-moments of the TMD PDFs, 
$
	f^{(n)}(x) = \int \dif{^2\kT}
					\left[ \kTsq / (2M^2) \right]^n f(x,\kTsq).
$
%\beq
%	\label{eq:wpdf}
%	f^{(n)}(x) = \int \dif{^2\kT}
%					\biggl( \frac{\kTsq}{2M^2} \biggr)^n f(x,\kTsq).
%\eeq
We measure the \qTscalt-weighted TSAs, defined as
\beq
	\label{eq:wasym}
	\A{T}{\sin{\Phi} \, W_{\Phi}}
			= \frac	{\int \dif{^2\qT} W_{\Phi} \, \F{T}{\sin\Phi}}
					{\int \dif{^2\qT} \F{U}{1}},
	\qquad
	\Phi = \phiS, 2\phi+\phiS, 2\phi-\phiS
\eeq
where $W_\Phi$ denotes the weights. The weighted TSAs are obtained by fit of the so-called modified double ratio 
$
	R(\Phi) \propto \A{T}{\sin\Phi W_{\Phi}} \, \sin\Phi.
$
The ratio is constructed from event counts and sums of event weights coming from two oppositely-polarised target cells and from two sub-periods divided by polarisation reversal to cancel the acceptance $a(\Phi)$. It is calculated using eight bins in $\Phi$ and one or three bins in four kinematic variables. As in the standard TSA analysis\,\cite{compass:2017dy} we use the dilution factor to correct the asymmetries for the target composition, so the TSAs refer to proton.

\begin{figure}
\includegraphics[width=0.95\textwidth]{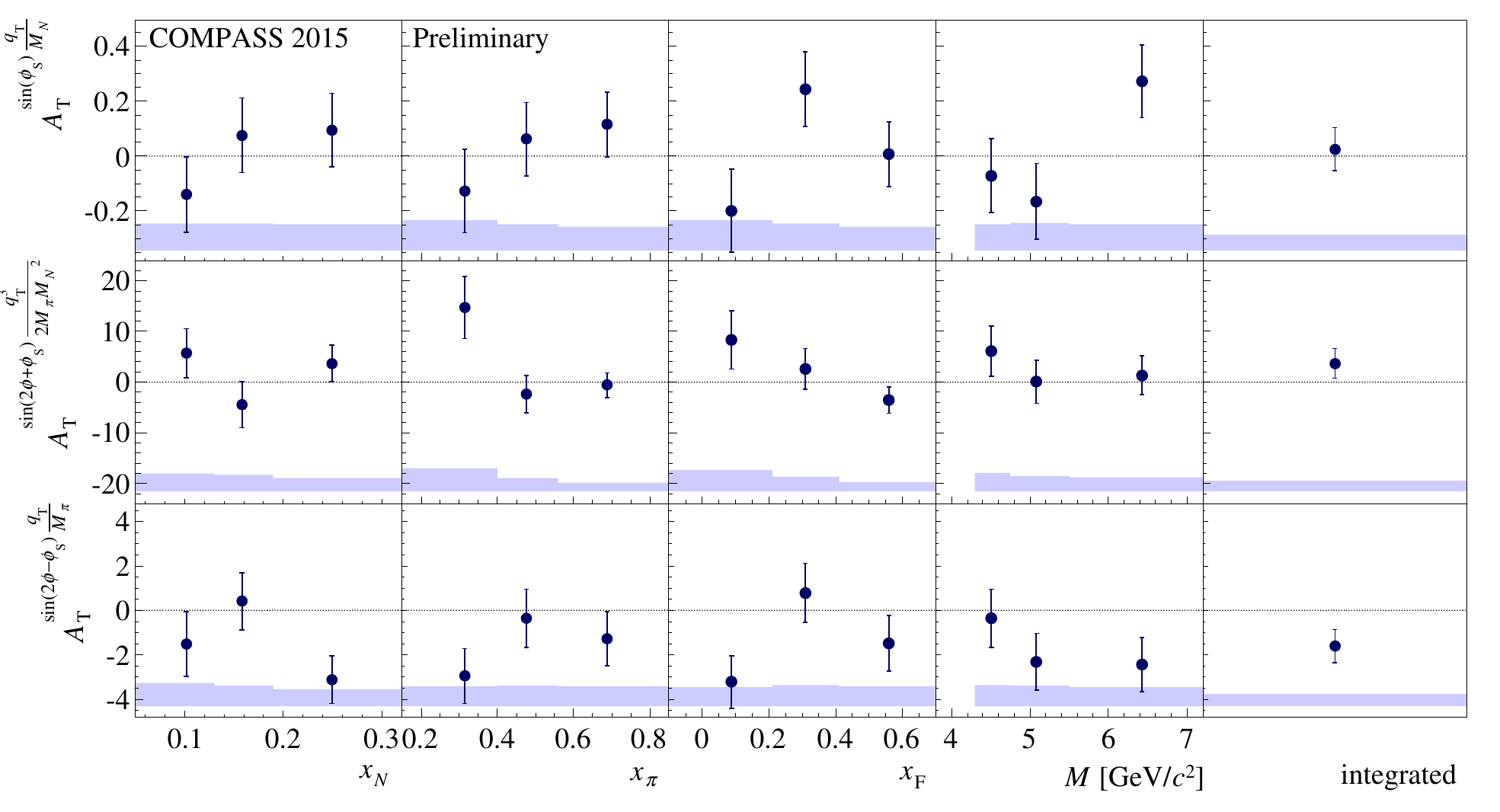}
\vspace{-10pt}
\caption{\label{fig:wasym} The \qTscalt-weighted TSAs. The systematic uncertainty is denoted by blue bands. Normalisation uncertainties of about 5\% (from polarisation and dilution factor) are not shown.}
\end{figure}

The data have been collected in 2015 in nine data-taking periods (each having two sub-periods). The event sample is almost the same as in the TSA analysis\,\cite{compass:2017dy}. The same invariant mass range $M\in[4.3, 8.5]$\,\GeVcc\ is used. The sharp cuts on \qTscalt\ are replaced by cut on individual muon transverse momenta $l_\mathrm{T} < 7$\,\GeVc. About 39\,000~dimuons pass the event selection. We estimate the background to be at the level of up to 4\,\%\,\cite{compass:2017dy}.
Several possible systematic effects have been investigated. The major contribution comes from the effect of variation of the data-taking conditions within a given period, estimated by measurement of false asymmetries. They are calculated from events with sub-periods or target cells of origin changed in such a way that the physics asymmetries cancel. The combined systematic uncertainty is about 0.7 times the statistical one. In addition, there are normalisation uncertainties of about 5\,\% from the polarisation measurement and dilution factor calculation. The results are shown on Fig.~\ref{fig:wasym}, the \qTscalt\ distribution on Fig.~\ref{fig:qT}.

%%%%%%%%%%%%%%%%%%%%%%%%%%%%%%%%%%%%%%%%%%%%%%%%%%%%%%%%%%%%%%%%%%%%%%%%%%%%%%%
\section{Transverse momentum weighted Sivers asymmetry in SIDIS and Drell--Yan}
\label{sec:projection}
%%%%%%%%%%%%%%%%%%%%%%%%%%%%%%%%%%%%%%%%%%%%%%%%%%%%%%%%%%%%%%%%%%%%%%%%%%%%%%%
To get a projection for the expected \qTscalt-weighted Sivers asymmetry in DY, we use the corresponding $\PhTscal/z$-weighted asymmetry in SIDIS $\mu \mathrm{p}^\uparrow \rarr \mu^\prime h X$, measured by COMPASS for positive and negative hadrons $h$ with $z>0.2$\,\cite{bradamante:2017}. It can be written as\,\cite{boer:1998,bradamante:2017}:
\beq
	\label{eq:wSivSIDISdef}
	\A{UT,T}{\sin(\phiH-\phiS) \frac{\PhTscal}{zM}} (x,z,Q^2)
			= \frac	{\int \dif{^2\PhT} \frac{\PhTscal}{zM}
						\, \F{UT,T}{\sin(\phiH-\phiS)}}
					{\int \dif{^2\PhT} \F{UU,T}{1}}
		= 2\frac{\sum_q e_q^2 \, f_{1\mathrm{T,p}}^{\perp (1) q} (x,Q^2) \,
								 D_{1,q}^h(z,Q^2)}
				 {\sum_q e_q^2 \, f_{1,\mathrm{p}}^q(x,Q^2) \, D_{1,q}^h(z,Q^2)},
\eeq
where we use the standard SIDIS variables, $F$ indicate the SIDIS structure functions\,\cite{bacchetta:2007}, and $D_{1,q}^h(z)$ is the fragmentation function of $q$ into hadron $h$. In writing explicitly the asymmetry, we only consider u, d, and s quarks and the corresponding antiquarks $q$ within the proton; we assume vanishing Sivers function of sea quarks; 
%Under these assumptions we have from~\refeq{eq:wSivSIDISdef}
%\beq
%	\label{eq:wSivSIDISfit}
%	A_{\mathrm{UT,T},h^\pm}^{\sin(\phiH-\phiS) \frac{\PhTscal}{zM}}(x,Q^2)
%		= 2\frac{ 	\frac{4}{9} \,	\fSiv{(1)\mathrm{u}}(x,Q^2) \, 
%									\tilde{D}_{1,\mathrm{u}}^{h^\pm}(Q^2)
%				  + \frac{1}{9} \,	\fSiv{(1)\mathrm{d}}(x,Q^2) \,
%									\tilde{D}_{1,\mathrm{d}}^{h^\pm}(Q^2)}
%				 {\sum_{q=\mathrm{u,\bar{u},d,\bar{d},s,\bar{s}}} 
%						e_q^2 \, f_1^q(x,Q^2) \tilde{D}_{1,q}^{h^\pm}(Q^2)},
%\eeq
%where $\tilde{D}_{1,q}^{h^\pm}$ are the fragmentation functions integrated over the available range in $z$.
and, like in Ref.\,\cite{martin:2017}, we integrate the fragmentation functions over the available range in $z$. 

\begin{figure}
\begin{minipage}{0.31\textwidth}
\includegraphics[width=0.8\textwidth]{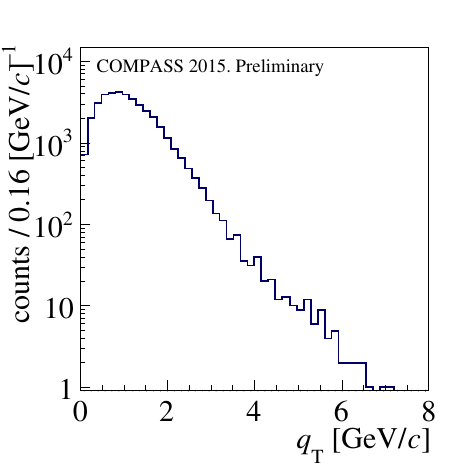}
\caption{\label{fig:qT} 
		Distribution of $\qTscal$ in the selected Drell--Yan events.}
\end{minipage}\hspace{0.025\textwidth}%
\begin{minipage}{0.31\textwidth}
\includegraphics[width=\textwidth]{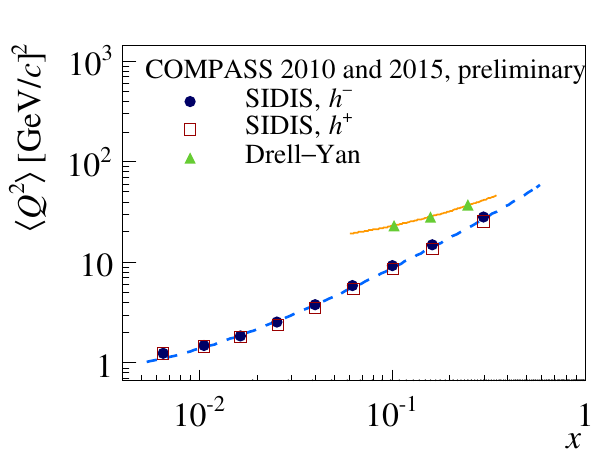}
\caption{\label{fig:Qsq}
	Mean $Q^2$ of the events in the Drell--Yan analysis
	and SIDIS analysis\,\cite{bradamante:2017}.}
\end{minipage}\hspace{0.025\textwidth}% 
\begin{minipage}{0.31\textwidth}
\includegraphics[width=\textwidth]{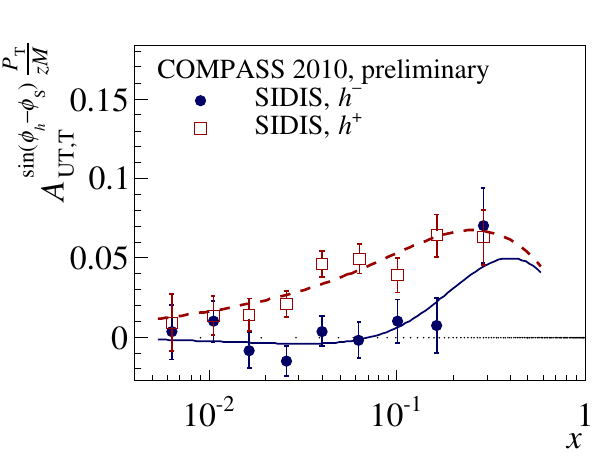}
\caption{\label{fig:AwSIDIS}
	The weighted Sivers asym. in SIDIS\,\cite{bradamante:2017}, fitted. 
	Statistical errors only.}
\end{minipage}
\end{figure}

We take the unpolarised PDFs from the CTEQ~5D global fit\,\cite{cteq5}, implemented in the LHAPDF library\,\cite{lhapdf6} and the charged hadron fragmentation functions from the DSS~07 LO global fit\,\cite{dss:2007}. As in\,\cite{martin:2017}, we use the collinear evolution of the PDFs and FFs, taken at the mean $Q^2$ at each $x$ (plotted on Fig.~\ref{fig:Qsq}), as $x$ and $Q^2$ are correlated. We parametrise the first $\kTsq$-moment of the Sivers function as
$
	\label{eq:SivParA}
	x \fSiv{(1)q}(x) = a_q \, x^{b_q} \, (1-x)^{c_q}.
$
The asymmetries for positive and negative hadrons in bins of $x$ are simultaneously fitted (Fig.~\ref{fig:AwSIDIS}).

\begin{figure}
\begin{minipage}{0.31\textwidth}
\includegraphics[width=\textwidth]{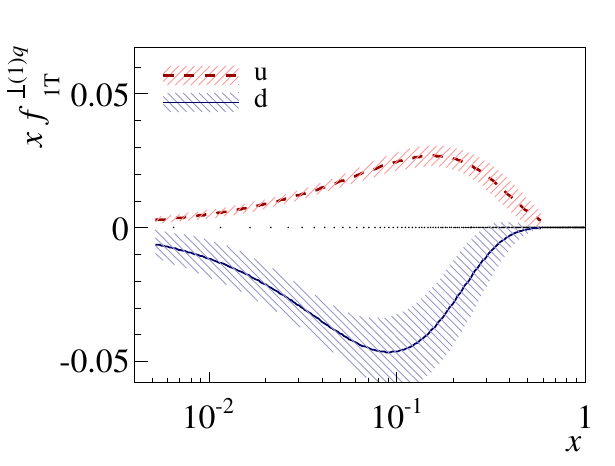}
\caption{\label{Sivers}
	The Sivers PDF first $\kTsq$-moment as a function of $x$ and 
	$Q^2(x)$. The $1\sigma$ error-bands account
	only for the uncertainty of the fit and the statistical errors 
	of the data.}
\end{minipage}\hspace{0.025\textwidth}%
\begin{minipage}{0.31\textwidth}
\includegraphics[width=\textwidth]{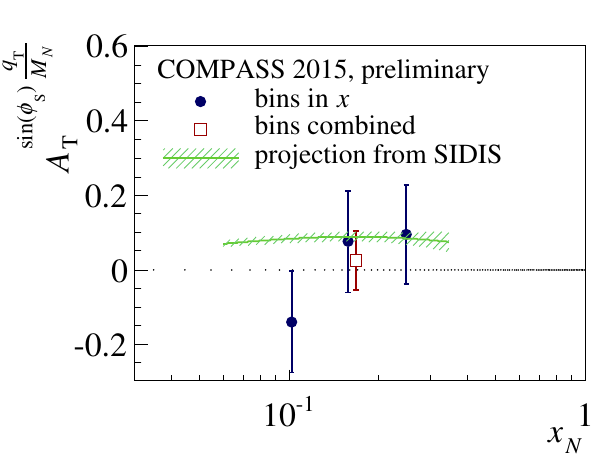}
\caption{\label{fig:AwDY}
	Weighted Sivers asymmetry in Drell--Yan from 2015 data and the 
	projection from corresponding asymmetry in SIDIS. 
	Only statistical errors are shown.}
\end{minipage}\hspace{0.025\textwidth}%
\begin{minipage}{0.31\textwidth}
\includegraphics[width=\textwidth]{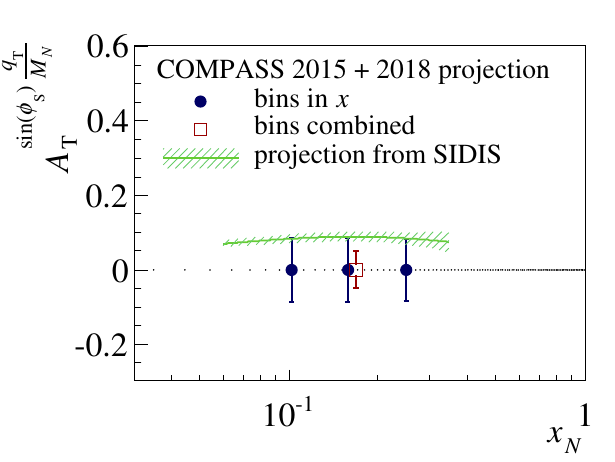}
\caption{\label{AwDYproj}
	A projection for combined analysis of 2015 and 2018 data. The 
	statistics in 2018 is assumed 1.5 times larger than in 2015.
	Only statistical errors are shown.}
\end{minipage} 
\end{figure}

To get the projection for the \qTscalt-weighted asymmetry in DY, we use Eq.~(\ref{eq:dy_ftphis_w}, \ref{eq:wasym}), we assume the change-of-sign prediction\,\cite{collins:2002} and valence quark dominance. The asymmetry simplifies to
$
	\A{T}{\sin\phiS \qTscal/M_\mathrm{p}}(x_N,Q^2)
		\approx 2  f_{1\mathrm{T,p}}^{\perp (1) \mathrm{u}}(x_N,Q^2)
				/ f_{1,\mathrm{p}}^\mathrm{u}(x_N,Q^2).
$
%\beq
%	\label{eq:SivDY}
%	\A{T}{\sin\phiS \frac{\qTscal}{M_\mathrm{p}}}(x_N,Q^2)
%		\approx 2  \frac{ f_{1\mathrm{T,p}}^{\perp (1) \mathrm{u}}(x_N,Q^2)}
%						{f_{1,\mathrm{p}}^\mathrm{u}(x_N,Q^2)}.
%\eeq
We identify the $x_N$ with the Bjorken $x$ from SIDIS and we use the same unpolarised PDF, taken at the mean $Q^2$ of the DY events used in the analysis (Fig.~\ref{fig:Qsq}). No evolution of $f_{1\mathrm{T,p}}^{\perp (1) \mathrm{u}}$ between the SIDIS and DY kinematics is considered. The result, compared with the measured asymmetries (Sec.~\ref{sec:measurement}), is shown on Fig.~\ref{fig:AwDY}. A projection for combined analysis of 2015 and 2018 data is shown as well, assuming the statistics in 2018 to be 1.5 times larger than in 2015. The $1\sigma$ error-bands account only for the uncertainty of the fit and the statistical errors of the experimental data. Variation of PDF and FF sets has been found to lead to differences of about 0.02. 

\ack
The author kindly acknowledges the support from Votruba--Blokhintsev programme and  Ministry of Education, Youth and Science of Czech Republic grant LM2015058.

\providecommand{\newblock}{}


\begin{thebibliography}{10}
\expandafter\ifx\csname url\endcsname\relax
  \def\url#1{{\tt #1}}\fi
\expandafter\ifx\csname urlprefix\endcsname\relax\def\urlprefix{URL }\fi
\providecommand{\eprint}[2][]{\url{#2}}

\bibitem{bacchetta:2007}
Bacchetta A {\em et~al.\/} 2007 {\em J. High Energy Phys.\/} {\bf 02} 093
  (\textit{Preprint}
  \href{http://arxiv.org/abs/hep-ph/0611265}{arXiv:hep-ph/0611265})

\bibitem{arnold:2008}
Arnold S, Metz A and Schlegel M 2009 {\em Phys. Rev.\/} {\bf D79} 034005
  (\textit{Preprint} \href{http://arxiv.org/abs/0809.2262}{arXiv:0809.2262
  [hep-ph]})

\bibitem{collins:2002}
Collins J~C 2002 {\em Phys. Lett.\/} {\bf B536} 43--48 (\textit{Preprint}
  \href{http://arxiv.org/abs/hep-ph/0204004}{arXiv:hep-ph/0204004})

\bibitem{compass:2017dy}
Aghasyan M {\em et~al.\/} (COMPASS) 2017 {\em Phys. Rev. Lett.\/} {\bf 119}
  112002 (\textit{Preprint}
  \href{http://arxiv.org/abs/1704.00488}{arXiv:1704.00488 [hep-ex]})

\bibitem{efremov:2004}
Efremov A~V {\em et~al.\/} 2005 {\em Phys. Lett.\/} {\bf B612} 233--244
  (\textit{Preprint}
  \href{http://arxiv.org/abs/hep-ph/0412353}{arXiv:hep-ph/0412353})

\bibitem{sissakian:2005b}
Sissakian A {\em et~al.\/} 2006 {\em Eur. Phys. J.\/} {\bf C46} 147--150
  (\textit{Preprint}
  \href{http://arxiv.org/abs/hep-ph/0512095}{arXiv:hep-ph/0512095})

\bibitem{kotzinian:1996}
Kotzinian A~M and Mulders P~J 1996 {\em Phys. Rev.\/} {\bf D54} 1229--1232
  (\textit{Preprint}
  \href{http://arxiv.org/abs/hep-ph/9511420}{arXiv:hep-ph/9511420})

\bibitem{kotzinian:1997}
Kotzinian A~M and Mulders P~J 1997 {\em Phys. Lett.\/} {\bf B406} 373--380
  (\textit{Preprint}
  \href{http://arxiv.org/abs/hep-ph/9701330}{arXiv:hep-ph/9701330})

\bibitem{boer:1998}
Boer D and Mulders P~J 1998 {\em Phys. Rev.\/} {\bf D57} 5780--5786
  (\textit{Preprint}
  \href{http://arxiv.org/abs/hep-ph/9711485}{arXiv:hep-ph/9711485})

\bibitem{gregor:2005}
Gregor I~M (HERMES) 2005 {\em Acta Phys. Polon.\/} {\bf B36} 209--215

\bibitem{bradamante:2017}
Bradamante F (COMPASS) 2017 {\em {Proceedings of the 22nd International Spin
  Symposium, Urbana-Champaign, USA, 25--30 September 2016}\/}
  (\textit{Preprint} \href{http://arxiv.org/abs/1702.00621}{arXiv:1702.00621
  [hep-ex]})

\bibitem{compass:prop2}
Gautheron F {\em et~al.\/} (COMPASS) 2010 {\em {COMPASS-II Proposal}\/}
  CERN-SPSC-2010-014, SPSC-P-340 (Geneva: CERN)
  \urlprefix\url{http://cds.cern.ch/record/1265628}

\bibitem{martin:2017}
Martin A, Bradamante F and Barone V 2017 {\em Phys. Rev.\/} {\bf D95} 094024
  (\textit{Preprint} \href{http://arxiv.org/abs/1701.08283}{arXiv:1701.08283
  [hep-ph]})

\bibitem{trento}
Bacchetta A {\em et~al.\/} 2004 {\em Phys. Rev.\/} {\bf D70} 117504
  (\textit{Preprint}
  \href{http://arxiv.org/abs/hep-ph/0410050}{arXiv:hep-ph/0410050})

\bibitem{cteq5}
Lai H~L {\em et~al.\/} (CTEQ) 2000 {\em Eur. Phys. J.\/} {\bf C12} 375--392
  (\textit{Preprint}
  \href{http://arxiv.org/abs/hep-ph/9903282}{arXiv:hep-ph/9903282})

\bibitem{lhapdf6}
Buckley A {\em et~al.\/} 2015 {\em Eur. Phys. J.\/} {\bf C75} 132
  (\textit{Preprint} \href{http://arxiv.org/abs/1412.7420}{arXiv:1412.7420
  [hep-ph]})

\bibitem{dss:2007}
de~Florian D, Sassot R and Stratmann M 2007 {\em Phys. Rev.\/} {\bf D75} 114010
  (\textit{Preprint}
  \href{http://arxiv.org/abs/hep-ph/0703242}{arXiv:hep-ph/0703242})

\end{thebibliography}
\end{document}